# On-Surface Synthesis of Graphene Nanoribbons on Two-Dimensional Rare Earth-Gold Intermetallic Compounds

Yande Que,[1#*] Bin Liu,[1#] Yuan Zhuang,[1] Chaoqiang Xu,[1] Kedong Wang,[2*] and Xudong Xiao[1*]

[1] *Department of Physics, the Chinese University of Hong Kong, Shatin, Hong Kong, China*

[2] *Department of Physics, Southern University of Science and Technology, Shenzhen, Guangdong 518055, P. R. China*

[#] *These authors contributed equally.*

[*] Corresponding author.

Email: ydque@phy.cuhk.edu.hk, wangkd@sustech.edu.cn, and xdxiao@phy.cuhk.edu.hk

**Abstract:**

Here, we demonstrate two reliable routes for the fabrication of armchair-edge graphene nanoribbons (GNRs) on $TbAu_2$/Au(111), belonging to a class of two-dimensional ferromagnetic rare earth-gold intermetallic compounds. On-surface synthesis directly on $TbAu_2$ leads to the formation of GNRs, which are short and interconnected with each other. In contrast, the intercalation approach –on-surface synthesis of GNRs directly on Au(111) followed by rare earth intercalation – yields GNRs on $TbAu_2$/Au(111), where both the ribbons and $TbAu_2$ are of high quality comparable with those directly grown on clean Au(111). Besides, the as-grown ribbons retain the same band gap while changing from p-doping to weak n-doping mainly due to a change in the work function of the substrate after the rare earth intercalation. The intercalation approach might also be employed to fabricate other types of GNRs on various rare earth intermetallic compounds, providing platforms to tailor the electronic and magnetic properties of GNRs on magnetic substrates.

**TOC Graphic:**

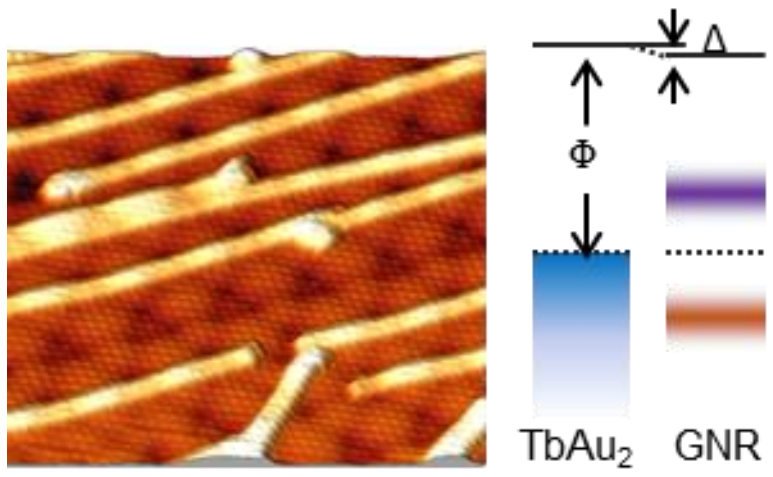

Graphene nanoribbons (GNRs), cut from graphene sheets, have received increasing interest among physicists and material scientists mainly due to the tunable electronic and magnetic properties as well as the great advances in the fabrication of GNRs with precise control.[1–5] For instance, bottom-up methods, like on-surface synthesis, have been employed to fabricate GNRs with precise edges and/or dopants using precursor molecules on various metal substrates in the past decade.[4,6–11] Substrate effects play a significant role in the on-surface synthesis of GNRs, e.g., determining the growth kinetics and even the final forms of the as-grown GNRs. Take the precursor of 10,10′-dibromo-9,9′-bianthryl (DBBA) as an example, it yields GNRs with an armchair edge and a width N of 7 carbon atoms (7-AGNRs) on the Au(111) surface via the standard two-step procedure.[1] On the contrary, only GNRs with chiral edges were formed on Cu(111) via a similar approach.[12,13] In addition, tuning the coupling between GNRs and their host substrate is an effective way to tailor the electronic and magnetic properties for specific applications. For instance, the band gap of GNRs can be tuned by changing the substrate from metals to insulators and thus varying the coupling with the substrate.[14–18] Further, extraordinary magnetisms emerge in GNRs with specific edges or termini.[19–21] which might be tuned by varying the hybridization between the π states in GNRs and the metal 3*d* or 4*f* orbitals of the ferromagnetic substrate. However, the previous works on GNRs mainly focused on the very limited metal substrates, like Au(111),[1,2,22–24] Au(110),[25] Ag(111),[26–28] and Cu(111),[12,13,29] which are paramagnetic (or nonmagnetic) and relatively inert. For other transition metal substrates, particularly a ferromagnetic metal substrate such as Fe, Co, and Ni, the strong interaction between the precursor overlayers with the metal substrate hinders the complex on-surface reactions, which are essential to the formation of GNRs.

On the other hand, two-dimensional (2D) ferromagnetic orders have been experimentally observed in $GdAu_2$/Au(111) and $GdAg_2$/Ag(111), where the magnetism mainly originates from the unpaired 4*f* electrons of Gd atoms.[30,31] These systems belong to the class of rare earth intermetallic compounds. Hence, the magnetism in such systems could be tuned by simply replacing the Gd atoms by other kinds of rare earth atoms with different 4*f* electrons,[32–34] while preserving similar physical (except magnetic) and chemical properties of these systems. Moreover, Abadía *et al.*[35] have demonstrated the polymerization of organic nanowires on the ferromagnetic $GdAu_2$/Au(111) substrate. In addition, their work reveals that the electronic states and the magnetic orders in the ferromagnetic substrate remain intact after the polymerization process. However, the polymerization process only involves the on-surface Ullmann reaction, the first step in the on-surface synthesis on GNRs. It is still unclear whether GNRs, *e.g.*, 7-AGNRs, could be synthesized on such a ferromagnetic substrate via on-surface synthesis requiring both the Ullmann reaction and the cyclodehydrogenation.[1,4]

In this Letter, we present a systematic study on the on-surface synthesis, structure, and electronic states of 7-AGNRs on the 2D rare earth-gold intermetallic compounds on Au(111) via low-temperature scanning tunneling microscopy/spectroscopy (LT-STM/STS). We first demonstrated that the on-surface synthesis of 7-AGNRs could occur on $TbAu_2$/Au(111). Contrary to the case on Au(111), most of the as-grown 7-AGNRs are short and interconnected with each other. Alternatively, we employed an intercalation approach to fabricate 7-GNRs on $TbAu_2$/Au(111), where both the 7-GNRs and $TbAu_2$ are of high quality comparable with those directly grown on clean Au(111) substrate. Moreover, the STS measurement was used to reveal the electronic properties of 7-GNRs/$TbAu_2$ systems.

Figure 1a shows a typical large-area STM image of the 2D rare earth intermetallic compound, $TbAu_2$ on Au(111) as the substrate for the on-surface synthesis of GNRs. It shows the well-ordered hexagonal moiré patterns with a periodicity of ~3.7 nm. The detailed growth and structures could be found elsewhere.[34] Deposition of the precursor DBBA molecules onto $TbAu_2$/Au(111) at room temperature gives rise to the formation of well-aligned molecular chains, as shown in Figure 1b. It shows three domains of these molecular chains with an orientation angle of 120˚ with respect to each other, indicating the chains follow the high-symmetry directions of the $TbAu_2$ lattice. Indeed, the orientation angle were measured to be 30˚ respective to the lattice vectors of the moiré structures as denoted by the black solid arrows (Figure 1b). Hence, these molecular chains are parallel to the lattice vectors of the $TbAu_2$ lattice, as the moiré lattice is rotated by 30˚ with respect to that of $TbAu_2$.[34] The inset in Figure 1b reveals two protrusions for each DBBA molecule, indicating the opposite tilt of the anthracene subunits in the DBBA molecules. The purple rectangle in the inset of Figure 1b marks the unit cell for the molecular chain structures with periodicities of 0.95 and 1.95 nm. Similar molecular chains have been observed for the DBBA molecules adsorbed on other metal substrates, such as Au(111)[36] (also see Figure S1 in the supporting information) and Cu(111).[12]

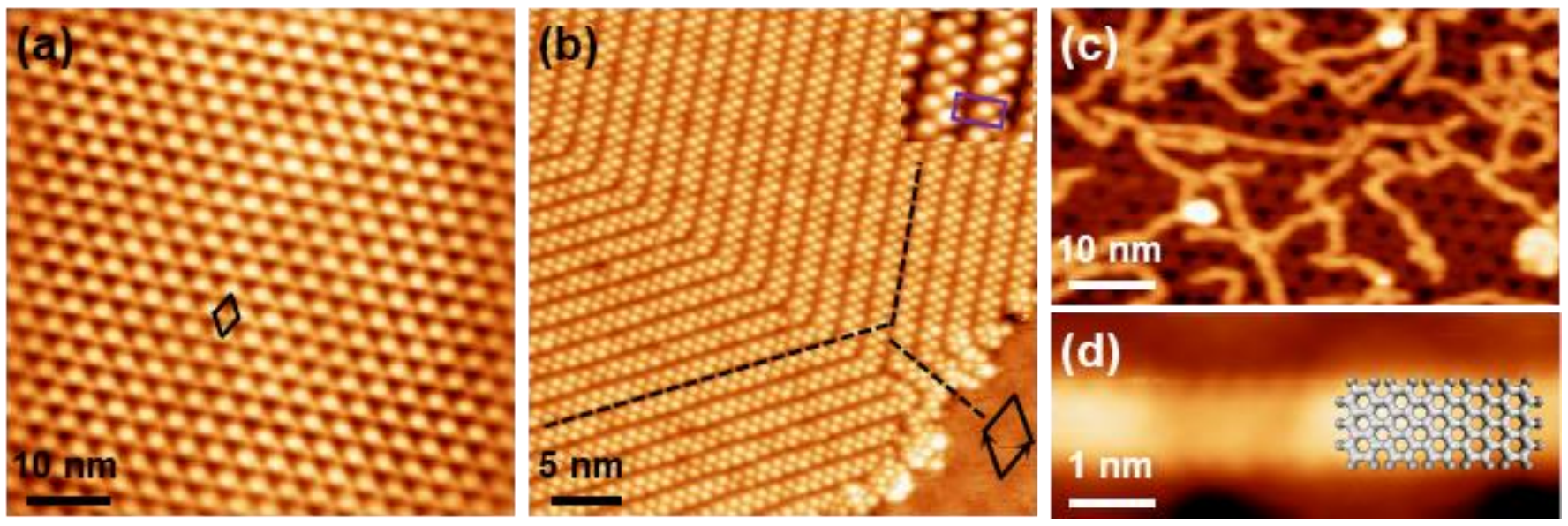

**Figure 1** On-surface synthesis of 7-AGNRs on $TbAu_2$/Au(111). (a) STM image of well-ordered $TbAu_2$ grown on Au(111). (b) STM image of DBBA molecular chains on $TbAu_2$/Au(111). Inset shows the zoomed-in image. (c) Large-area and (d) atomic-resolution STM images of 7-AGNRs on $TbAu_2$/Au(111). Tunneling parameters: (a-c) sample bias $V_S$ = 2.0 V, tunneling current $I_T$ = 0.2 nA; (d) $V_S$ = 0.1 V, $I_T$ = 0.5 nA.

A standard two-step annealing procedure was employed to stimulate the formation of GNRs on $TbAu_2$/Au(111). Figure 1c presents the typical STM image of the as-grown 7-AGNRs on $TbAu_2$/Au(111). It shows these ribbons with roughly identical width but with different lengths ranging from a few nanometers to a few tens of nanometers. As shown in Figure 1d, the armchair edges were revealed by the atomically resolved STM image, which is in good agreement with the 7-AGNRs model. Although most of these ribbons are randomly aligned on the surface and interconnected with each other, it implies the successful on-surface synthesis of 7-AGNRs on the $TbAu_2$/Au(111).

The substrate and annealing procedure play an essential role in the on-surface synthesis of GNRs of high quality.[37–39] Further optimizing the two-step annealing process might improve the quality of the as-grown 7-AGNRs on $TbAu_2$/Au(111) by the on-surface synthesis. Alternatively, we employed an intercalation approach to fabricate high-quality 7-AGNRs on $TbAu_2$/Au(111), which has been widely used to tune the coupling of the graphene and the layer beneath.[40–44] Briefly, high-quality 7-AGNRs were first grown on Au(111) via the on-surface synthesis. Subsequently, the well-ordered $TbAu_2$ layer was achieved by depositing Tb atoms onto the hot (300 ˚C) substrate of 7-AGNRs/Au(111), as sketched in Figure 2a.

Figure 2b shows the high-quality 7-AGNRs on Au(111) via on-surface synthesis. It reveals the straight, long (up to 60 nm) ribbons, mainly following the high-symmetry direction of the Au(111) surface, *e.g.*, parallel to the herringbones or 30˚ respective to the herringbones of Au(111). As shown in Figure 2c, the high-resolution STM image reveals the periodical edge patterns with a periodicity of 0.42 nm, indicating the armchair edges of the ribbon.[1]

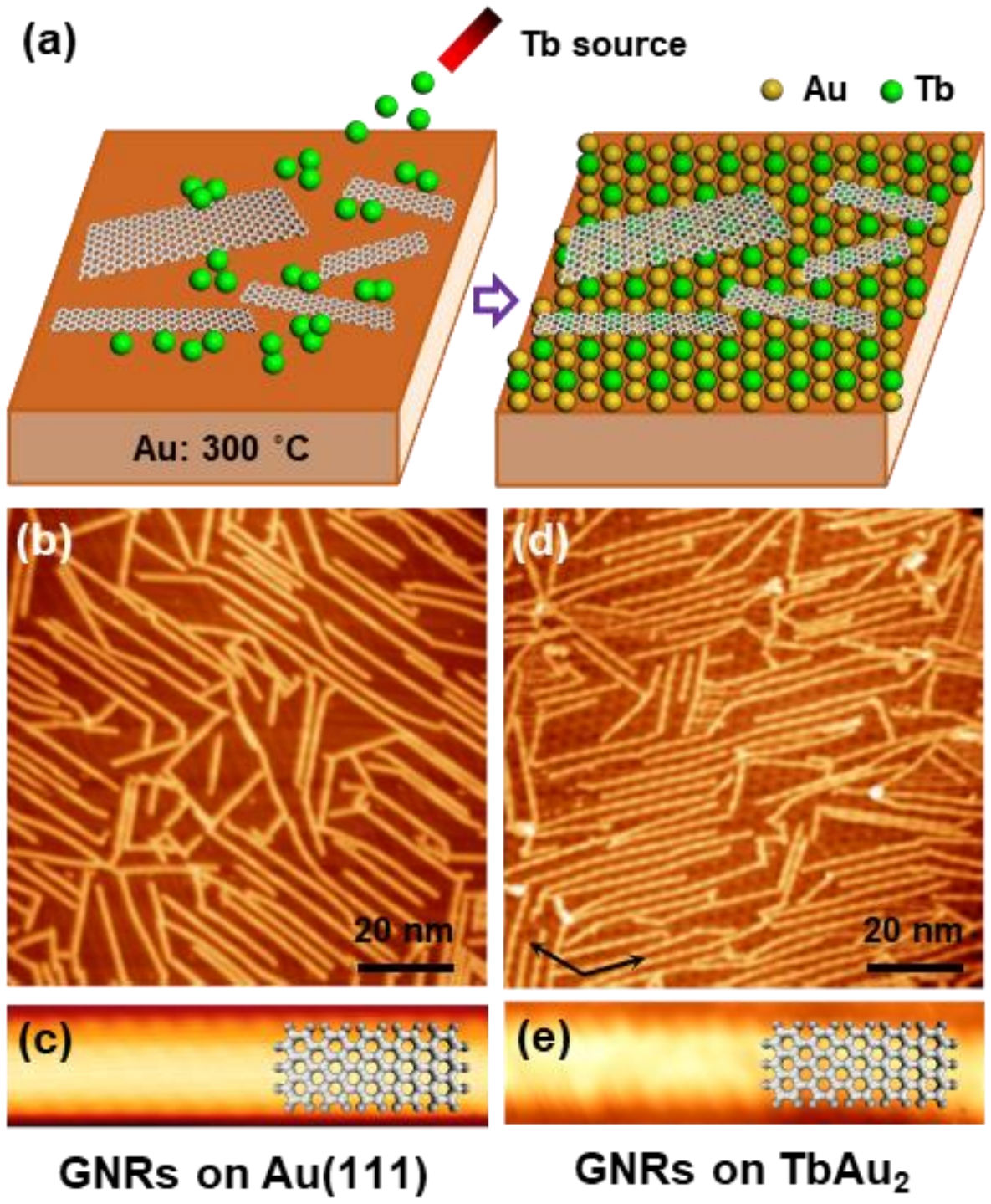

**Figure 2** Tb intercalation between armchair graphene nanoribbons and Au(111). (a) Schematic process of Tb interaction. Large-area and atomic-resolution STM topographic images of graphene nanoribbons on Au(111) before (b, c) and after (d, e) the Tb intercalation, respectively. A segment of the 7-AGNR model is overlapped with the edge-resolved GNR in (c) and (e). The solid arrows denote the lattice directions of the moiré structure of $TbAu_2$. Tunneling parameters: (b) $V_S$ = -2.0 V, $I_T$ = 0.2 nA, (c) $V_S$ = 10 mV, $I_T$ = 1.0 nA, (d) $V_S$ = 2.0 V, $I_T$ = 0.2 nA, (e) $V_S$ = 10 mV, $I_T$ = 2.0 nA.

After the Tb intercalation, the 7-AGNRs remain straight and long comparable with those before the intercalation, implying a nondestructive intercalation process, as shown in Figure 2d. Contrary to the flat surface of the ribbons before the intercalation, it shows the periodically modulated surface of the ribbons by the long-range ordered moiré patterns of $TbAu_2$/Au(111). Besides, it reveals that the ribbons are mainly aligned with the high-symmetry directions of the substrate, *e.g.* along the direction of or 30$^\circ$ with respect to the moiré pattern of $TbAu_2$/Au(111), contrary to the random alignment of those directly on-surface synthesized on the alloy substrate (Figure 1c). The atomically resolved STM image in Fig. 2(e) illustrates the atomically precise edges without defects or impurities, indicating that the ribbons remain intact during the intercalation processes. In addition, it is worth noting that the well-ordered moiré pattern of the $TbAu_2$/Au(111) beneath the ribbons implying the existence of the ribbon has little effect to the formation of $TbAu_2$/Au(111) (also see Figure S2 in the supporting information). Moreover, 7-AGNRs on other rare earth-gold intermetallic compounds could be achieved by such intercalation approach. For instance, the deposition of another type of rare earth atom, Ho, onto the sample of 7-AGNRs/Au(111) leads to the formation of $HoAu_2$/Au(111) beneath the ribbons (see Figure S3 in the supporting information).

So far, we have demonstrated a reliable approach to fabricate high-quality 7-AGNRs on rare earth – gold intermetallic compounds combined on-surface synthesis of GNRs and rare earth intercalations. In the following, the STS measurement was employed to illustrate the electronic properties of 7-AGNRs on $TbAu_2$/Au(111).

Differential conductance (d*I*/d*V*) spectra of 7-AGNRs/Au were also taken for comparison, as shown in Figure 3a. It reveals the onsets of the valence band (VB) and conduction band (CB) at -1.0 eV and 1.7 eV, respectively. Thus, the band gap of 7-AGNRs/Au was measured to be 2.7 eV, which is in consistent with previous works.[45–47] In addition, the asymmetry in the VB and CB respective to the Fermi level ($E_F$) implies the p-doping effect on the ribbons due to the coupling with the Au substrate. On 7-AGNRs/$TbAu_2$, the dI/dV spectra in Figure 3b show the onsets of the VB and CB shift to -1.4 eV and 1.3 eV, respectively. As a result, the band gap of 7-AGNRs remains unchanged (2.7 eV) compared with that on Au(111). The small asymmetry in the VB and CB with respect to $E_F$ implies that the doping effect on the ribbons is significantly suppressed compared with that on Au(111). Moreover, the reversed symmetry reveals the doping effect to the ribbons changes from p-doping to n-doping compared with that on Au(111). Besides, the different shifts in the surface states (SS) of Au(111) and $TbAu_2$ with the existence of the ribbons should be noted. For the Shockley SS of Au(111), it shifts from -0.5 eV for bare Au(111) to -0.3 eV for GNRs/Au(111), whereas the $TbAu_2$ state around 0.6 eV shifts from 625 meV for bare $TbAu_2$ to 612 meV for GNRs/$TbAu_2$. Such differences in the surface states shift are in line with the change of the doping effects on the ribbons.

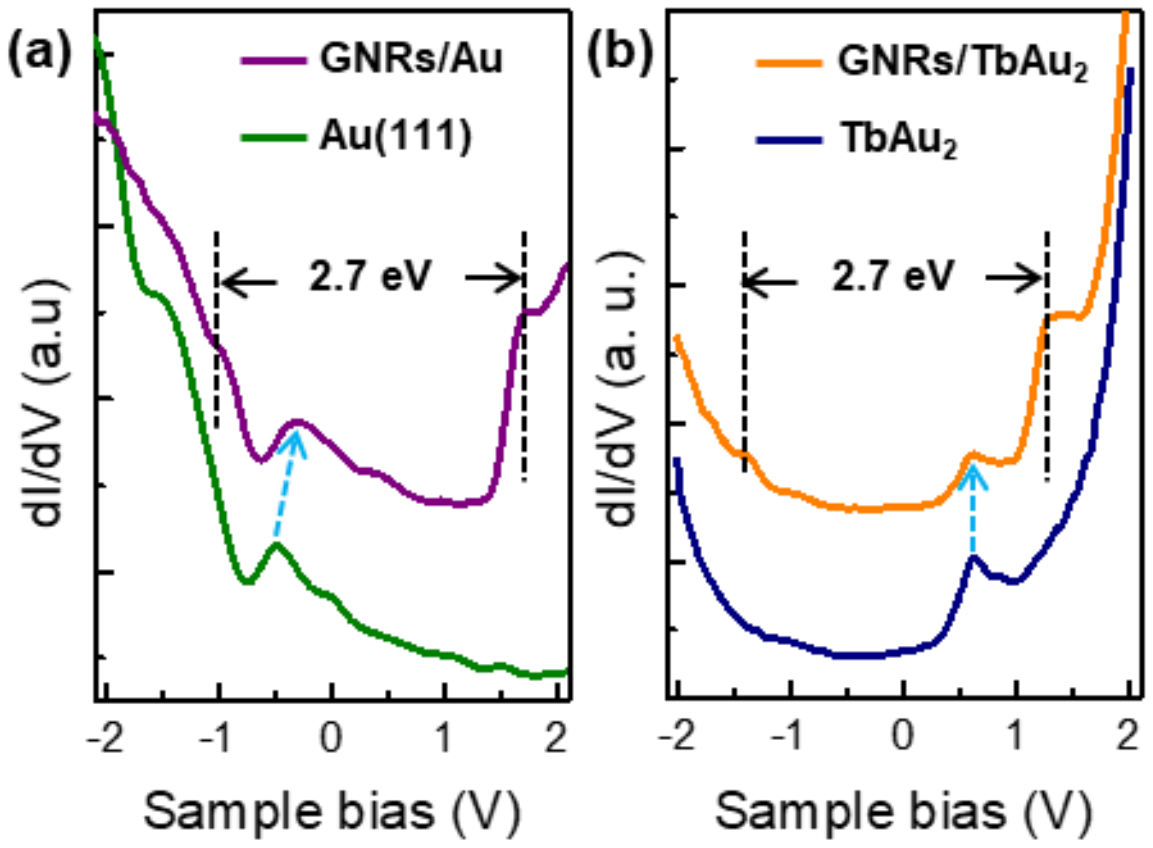

**Figure 3** Differential conductance spectra of 7-AGNRs (a) on Au(111) and (b) on $TbAu_2$/Au(111). Black dashed lines indicate the onsets of the VB and/or CB of 7-AGNRs. The values in between represent the corresponding band gap of the ribbons. Light blue arrow lines indicate the shift of the SS of Au(111) and/or $TbAu_2$. Tunneling parameters for the spectra (a, b): $V_S$ = 2.0 V, $I_T$ = 0.2 nA.

The band gap (2.7 eV) of 7-AGNRs on Au and $TbAu_2$ is smaller than that of decoupled 7-AGNRs (2.9~3.5 eV),[14,48] caused by the screening effect of the metallic substrate.[16] Besides, the metal metal-organic interfaces play an essential role in the electronic properties of the ribbons, like doping. The change in the work function has been used to depict the physical origins of the metal-organic interfaces.[49,50] To unveil the origins of the changes in the doping effect on the ribbons, we carried out the I(z) spectroscopy measurement to extract the local surface work functions (for more details, see S4 and Figure S4 in the supporting information). The I(z) spectra could be described by the following equation:[51]

$$I\left(z\right) \propto \exp\left(-2z\sqrt{2m\Phi_{ap}/\hbar^2}\right) \tag{1}$$

where $z$ is the STM tip-sample distance, $m$ is the mass of the tunneling electrons, $\Phi_{ap}$ is the apparent tunneling barrier, and $\hbar$ is the reduced Planck constant. As a rough estimation, the tunneling barrier could be approximated by $\Phi_{ap} = \left(\Phi_{tip} + \Phi_{sample}\right)/2$,[51,52] where $\Phi_{tip/sample}$ is the local work function of the tip or sample. Despite the fact that the work function of the STM tip is unknown and might be varied from time to time, the differences in the local surface work function over the sample could be extracted using the same STM tip, namely, $\Delta\Phi = 2\times\left(\Phi_{ap}^{GNR} - \Phi_{ap}^{sub}\right)$. It should be noted that the field emission resonant (FER) states could also be employed to measure the local surface work functions.[34,53] However, the high tip-induced electric field during the measurement might induce the ribbons unstable on the surface (see S5 and Figure S5 in the supporting information), giving rise to high uncertainty in such a measurement.

Figure 4a,b presents the I(z) spectra taken on 7-AGNRs/Au(111) and 7-AGNRs/$TbAu_2$ as well as their substrates, respectively. All these curves could be well fitted by the equation (1), giving rise to the different apparent tunneling barriers, as shown in Figure 4c. On Au(111), it reveals a reduction of ~0.9 eV in the apparent tunneling barrier with the existence of 7-AGNRs (Au(111): 4.97 ± 0.15 eV, 7-AGNRs/Au: 4.09 ± 0.26 eV), indicating that the local surface work function of 7-AGNRs/Au is lower than that of bare Au(111) by 1.8 eV. Such a change in the local surface work function can be explained by the formation of the interface dipole between the Au substrate and the ribbons. Figure 5a illustrates that a surface dipole with potential of Δ (equal to the local surface work function reduction of 1.8 eV) is formed as the ribbon contacts with the Au substrate

due to the differences in their work functions (~5.3 eV for Au(111)[34,54] and ~4.5 eV for graphene-related materials[55]). Consequently, the surface dipole field results in the p-doping to the ribbons.

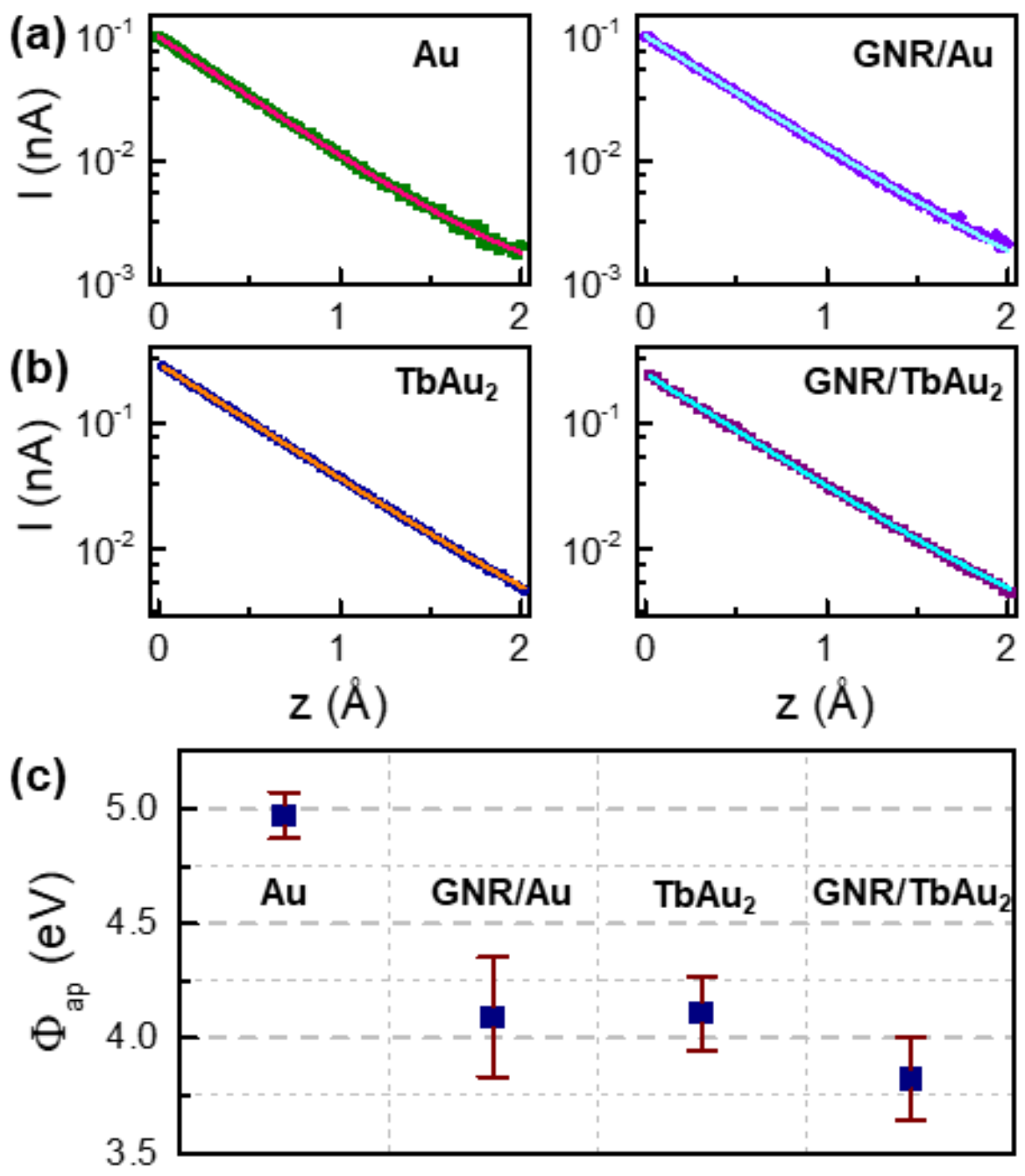

**Figure 4** I(z) spectra taken (a) on Au(111) and 7-AGNR/Au, and (b) on $TbAu_2$/Au(111) and 7-AGNR/$TbAu_2$, respectively. The solid lines are fitted curves according to the equation (1). (c) Statistic results of the tunneling barrier extracted from the fittings of 8 spectra for each sample. The spectra in (a) and (b) were taken at $V_S$ = 2.0 V.

Similarly, it shows the same trend for the apparent tunneling barrier on $TbAu_2$/Au(111), namely, the apparent tunneling barrier slightly decreases by 0.2 eV from 4.02 ± 0.16 eV for bare $TbAu_2$ to 3.82 ± 0.18 eV for 7-AGNR/$TbAu_2$. As a result, the local surface work function of $TbAu_2$/Au(111) slightly decreases by 0.4 eV with the existence of the ribbons, much smaller compared with the case on Au(111). Such slight change in the local surface work function is consistent with the similar work functions for graphene-related materials and the rare earth – gold surface alloys.[34,55] Therefore, a weak interface dipole is formed between the ribbon and $TbAu_2$ substrate, as shown in Figure 5b. Compared with the case on Au(111), this weak interface dipole field would give rise to a weak p-doping to the ribbons, in contrast to the results of the dI/dV. Therefore, other effects should be taken into account to explain the weak n-doping in 7-AGNR/$TbAu_2$ systems, *e.g.* charge transfer[56] or hybridization between the ribbons and $TbAu_2$ substrate. More experimental and theoretic works are required to unravel the origins of such differences.

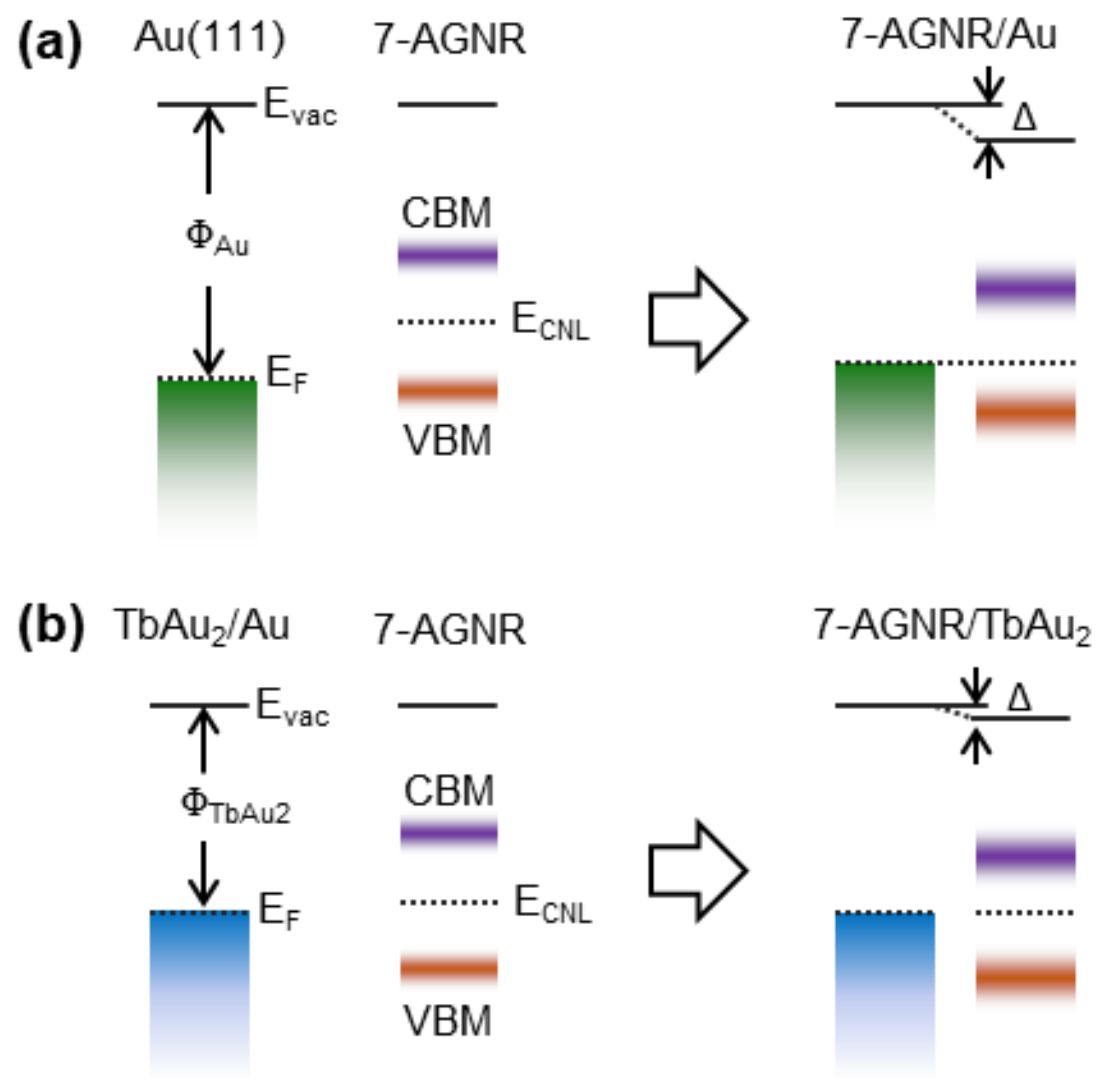

**Figure 5** Energy band diagram for 7-AGNRs on (a) Au(111) and (b) $TbAu_2$/Au(111). $E_{vac}$ represents the vacuum energy level; Φ represents the work function; $E_{CNL}$ represents the charge neutral energy level for the ribbons; VBM represents the valence band maximum and CBM represents the conduction band minimum.

In conclusion, we have demonstrated two reliable routes for fabrication of 7-AGNRs on 2D rare earth-gold intermetallic compound $TbAu_2$/Au(111). The as-grown ribbons via on-surface synthesis directly on $TbAu_2$ are short and interconnected with each other. In contrast, the intercalation approach – on-surface synthesis of 7-AGNRs directly on Au(111) followed by rare earth intercalation – yields 7-AGNRs on $TbAu_2$/Au(111), where both the ribbons and $TbAu_2$ are of high quality comparable with those directly grown on Au(111). Besides, the STS measurement reveals that the doping effects on the ribbons change from p-doping to weak n-doping due to the different work functions of the substrate while the band gap remains unchanged after the rare earth intercalation. The intercalation approach might also be employed to fabricate other types of GNRs on various rare earth intermetallic compounds, providing platforms to tailor the electronic and magnetic properties of GNRs on magnetic substrates.

## Materials and Methods

*Sample preparation.* All the samples were prepared in an ultrahigh vacuum (base pressure of ~1 × $10^{-10}$ millibar) LT-STM system. Single crystal Au(111) (MaTeck) was cleaned by several cycles of ion sputtering using $Ar^+$ with an energy of 600 eV and annealing at 450 ˚C until STM topographic images show well-defined reconstructed herringbone structures of Au(111) and free of surface impurities. 7-AGNRs on Au(111) [$TbAu_2$/Au(111)] were grown via the on-surface synthesis method.[1] Briefly, a sub-monolayer of DBBA chains was firstly prepared by depositing DBBA molecules onto the Au(111) [$TbAu_2$/Au(111)] surface (room temperature) using a homemade Knudsen cell at 135 ˚C. A subsequent two-step annealing (200 ˚C and 400 ˚C, heating and cooling rates of 30 ˚C/min) gives rise to the formation of 7-AGNRs. Further intercalation of $TbAu_2$ and/or $HoAu_2$ between GRNs and Au(111) was achieved by deposition of rare earth (Tb or Ho) atoms onto hot (300 ˚C) substrate of GNRs/Au, similar to the growth of these rare earth intermetallic compounds directly on Au(111).[34] The deposition rate was adjusted to around 0.13 monolayer $TbAu_2$ (evaporator power: 8 mA, 1 kV, Tb rod with diameter of 2 mm) or $HoAu_2$ (evaporator power: ~7 mA, 1 kV, Ho rod with diameter of 2 mm) per minute.

*STM/STS measurement.* STM images were acquired in a constant-current mode and all given voltages refer to the sample.[34,57] Numerical differentiation of I(V) spectra was employed to obtain the differential conductance (dI/dV) spectra (For more details, see S6 and Figure S6 in the supporting information). STM/STS experiments were carried out at ~78 K, using an electrochemically etched tungsten tip that was calibrated against the surface states of the Au(111) surface prior to and after spectroscopic measurements.

## ASSOCIATED CONTENT

### Supporting Information

The Supporting Information is available free of charge at https://pubs.acs.org/doi/10.1021/acs.jpclett.0c0139.

DBBA chains on Au(111), Additional STM image for GNRs on $TbAu_2$/Au(111) via intercalation approach, GNRs on $HoAu_2$/Au(111) via intercalation, Relationship between the Tunneling barrier and $I_0$ , Unstable ribbons under FER measurement, and Numerical differentiation of I(V) spectra (PDF)

## AUTHOR INFORMATION

### Corresponding Authors

*Email: ydque@phy.cuhk.edu.hk (Y.D.Q.).

*Email: wangkd@sustech.edu.cn (K. D. W.).

*Email: xdxiao@phy.cuhk.edu.hk (X.D.X.).

### ORCID

Yande Que: 0000-0002-5267-4985

Chaoqiang Xu: 0000-0001-8561-8974

Yuan Zhuang: 0000-0003-3634-8546

Xudong Xiao: 0000-0003-0551-1144

### Author Contributions

Y.D.Q. and B.L. Contributed equally to this work.

**Notes**

The authors declare no competing financial interest.

## ACKNOWLEDGEMENTS

We acknowledge the financial support from the RGC of Hong Kong (No. 404613), the Direct Grant for Research of CUHK (No. 4053306 and No. 4053348), NSFC (No. 11574128) and MOST (No. 2014CB921402).

## REFERENCES

(1) Cai, J.; Ruffieux, P.; Jaafar, R.; Bieri, M.; Braun, T.; Blankenburg, S.; Muoth, M.; Seitsonen, A. P.; Saleh, M.; Feng, X.; et al. Atomically Precise Bottom-up Fabrication of Graphene Nanoribbons. *Nature* **2010**, *466*, 470–473.

(2) Cai, J.; Pignedoli, C. A.; Talirz, L.; Ruffieux, P.; Söde, H.; Liang, L.; Meunier, V.; Berger, R.; Li, R.; Feng, X.; et al. Graphene Nanoribbon Heterojunctions. *Nat. Nanotechnol.* **2014**, *9*, 896–900.

(3) Ruffieux, P.; Wang, S.; Yang, B.; Sanchez-Sanchez, C.; Liu, J.; Dienel, T.; Talirz, L.; Shinde, P.; Pignedoli, C. A.; Passerone, D.; et al. On-Surface Synthesis of Graphene Nanoribbons with Zigzag Edge Topology. *Nature* **2016**, *531*, 489–492.

(4) Talirz, L.; Ruffieux, P.; Fasel, R. On-Surface Synthesis of Atomically Precise Graphene Nanoribbons. *Adv. Mater.* **2016**, *28*, 6222–6231.

(5) Li, J.; Sanz, S.; Corso, M.; Choi, D. J.; Peña, D.; Frederiksen, T.; Pascual, J. I. Single Spin Localization and Manipulation in Graphene Open-Shell Nanostructures. *Nat. Commun.* **2019**, *10*, 200.

(6) Chen, Z.; Wang, H. I.; Bilbao, N.; Teyssandier, J.; Prechtl, T.; Cavani, N.; Tries, A.; Biagi, R.; De Renzi, V.; Feng, X.; et al. Lateral Fusion of Chemical Vapor Deposited N = 5 Armchair Graphene Nanoribbons. *J. Am. Chem. Soc.* **2017**, *139*, 9483–9486.

(7) Cao, Y.; Qi, J.; Zhang, Y. F.; Huang, L.; Zheng, Q.; Lin, X.; Cheng, Z.; Zhang, Y. Y.; Feng, X.; Du, S.; et al. Tuning the Morphology of Chevron-Type Graphene Nanoribbons by Choice of Annealing Temperature. *Nano Res.* **2018**, *11*, 6190–6196.

(8) Gröning, O.; Wang, S.; Yao, X.; Pignedoli, C. A.; Borin Barin, G.; Daniels, C.; Cupo, A.;

Meunier, V.; Feng, X.; Narita, A.; et al. Engineering of Robust Topological Quantum Phases in Graphene Nanoribbons. *Nature* **2018**, *560*, 209–213.

(9) Rizzo, D. J.; Veber, G.; Cao, T.; Bronner, C.; Chen, T.; Zhao, F.; Rodriguez, H.; Louie, S. G.; Crommie, M. F.; Fischer, F. R. Topological Band Engineering of Graphene Nanoribbons. *Nature* **2018**, *560*, 204–208.

(10) Zhang, Y.-F.; Zhang, Y.; Li, G.; Lu, J.; Que, Y.; Chen, H.; Berger, R.; Feng, X.; Müllen, K.; Lin, X.; et al. Sulfur-Doped Graphene Nanoribbons with a Sequence of Distinct Band Gaps. *Nano Res.* **2017**, *10*, 3377–3384.

(11) Pedramrazi, Z.; Chen, C.; Zhao, F.; Cao, T.; Nguyen, G. D.; Omrani, A. A.; Tsai, H. Z.; Cloke, R. R.; Marangoni, T.; Rizzo, D. J.; et al. Concentration Dependence of Dopant Electronic Structure in Bottom-up Graphene Nanoribbons. *Nano Lett.* **2018**, *18*, 3550–3556.

(12) Han, P.; Akagi, K.; Federici Canova, F.; Mutoh, H.; Shiraki, S.; Iwaya, K.; Weiss, P. S.; Asao, N.; Hitosugi, T. Bottom-up Graphene-Nanoribbon Fabrication Reveals Chiral Edges and Enantioselectivity. *ACS Nano* **2014**, *8*, 9181–9187.

(13) Sánchez-Sánchez, C.; Dienel, T.; Deniz, O.; Ruffieux, P.; Berger, R.; Feng, X.; Müllen, K.; Fasel, R. Purely Armchair or Partially Chiral: Noncontact Atomic Force Microscopy Characterization of Dibromo-Bianthryl-Based Graphene Nanoribbons Grown on Cu(111). *ACS Nano* **2016**, *10*, 8006–8011.

(14) Wang, S.; Talirz, L.; Pignedoli, C. A.; Feng, X.; Müllen, K.; Fasel, R.; Ruffieux, P. Giant Edge State Splitting at Atomically Precise Graphene Zigzag Edges. *Nat. Commun.* **2016**, *7*, 11507.

(15) Simonov, K. A.; Vinogradov, N. A.; Vinogradov, A. S.; Generalov, A. V.; Svirskiy, G. I.; Cafolla, A. A.; Mårtensson, N.; Preobrajenski, A. B. Effect of Electron Injection in Copper-Contacted Graphene Nanoribbons. *Nano Res.* **2016**, *9*, 2735–2746.

(16) Deniz, O.; Sánchez-Sánchez, C.; Dumslaff, T.; Feng, X.; Narita, A.; Müllen, K.; Kharche, N.; Meunier, V.; Fasel, R.; Ruffieux, P. Revealing the Electronic Structure of Silicon Intercalated Armchair Graphene Nanoribbons by Scanning Tunneling Spectroscopy. *Nano Lett.* **2017**, *17*, 2197–2203.

(17) Wang, S.; Kharche, N.; Costa Girão, E.; Feng, X.; Müllen, K.; Meunier, V.; Fasel, R.; Ruffieux, P. Quantum Dots in Graphene Nanoribbons. *Nano Lett.* **2017**, *17*, 4277–4283.

(18) Jacobse, P. H.; Mangnus, M. J. J.; Zevenhuizen, S. J. M.; Swart, I. Mapping the Conductance of Electronically Decoupled Graphene Nanoribbons. *ACS Nano* **2018**, *12*, 7048–7056.

(19) Kunstmann, J.; Özdoğan, C.; Quandt, A.; Fehske, H. Stability of Edge States and Edge Magnetism in Graphene Nanoribbons. *Phys. Rev. B* **2011**, *83*, 045414.

(20) Magda, G. Z.; Jin, X.; Hagymási, I.; Vancsó, P.; Osváth, Z.; Nemes-Incze, P.; Hwang, C.; Biró, L. P.; Tapasztó, L. Room-Temperature Magnetic Order on Zigzag Edges of Narrow Graphene Nanoribbons. *Nature* **2014**, *514*, 608–611.

(21) Lawrence, J.; Brandimarte, P.; Berdonces-Layunta, A.; Mohammed, M. S. G.; Grewal, A.;

Leon, C. C.; Sánchez-Portal, D.; De Oteyza, D. G. Probing the Magnetism of Topological End States in 5-Armchair Graphene Nanoribbons. *ACS Nano* **2020**.

(22) Talirz, L.; Söde, H.; Dumslaff, T.; Wang, S.; Sanchez-Valencia, J. R.; Liu, J.; Shinde, P.; Pignedoli, C. A.; Liang, L.; Meunier, V.; et al. On-Surface Synthesis and Characterization of 9-Atom Wide Armchair Graphene Nanoribbons. *ACS Nano* **2017**, *11*, 1380–1388.

(23) Merino-Díez, N.; Li, J.; Garcia-Lekue, A.; Vasseur, G.; Vilas-Varela, M.; Carbonell-Sanromà, E.; Corso, M.; Ortega, J. E.; Peña, D.; Pascual, J. I.; et al. Unraveling the Electronic Structure of Narrow Atomically Precise Chiral Graphene Nanoribbons. *J. Phys. Chem. Lett.* **2018**, *9*, 25–30.

(24) Liu, M.; Chen, S.; Li, T.; Wang, J.; Zhong, D. Tuning On-Surface Synthesis of Graphene Nanoribbons by Noncovalent Intermolecular Interactions. *J. Phys. Chem. C* **2018**, *122*, 24415–24420.

(25) Massimi, L.; Ourdjini, O.; Lafferentz, L.; Koch, M.; Grill, L.; Cavaliere, E.; Gavioli, L.; Cardoso, C.; Prezzi, D.; Molinari, E.; et al. Surface-Assisted Reactions toward Formation of Graphene Nanoribbons on Au(110) Surface. *J. Phys. Chem. C* **2015**, *119*, 2427–2437.

(26) Huang, H.; Wei, D.; Sun, J.; Wong, S. L.; Feng, Y. P.; Neto, A. H. C.; Wee, A. T. S. Spatially Resolved Electronic Structures of Atomically Precise Armchair Graphene Nanoribbons. *Sci. Rep.* **2012**, *2*, 983.

(27) Simonov, K. A.; Generalov, A. V.; Vinogradov, A. S.; Svirskiy, G. I.; Cafolla, A. A.; McGuinness, C.; Taketsugu, T.; Lyalin, A.; Mårtensson, N.; Preobrajenski, A. B. Synthesis of Armchair Graphene Nanoribbons from the 10,10′-Dibromo-9,9′-Bianthracene Molecules on Ag(111): The Role of Organometallic Intermediates. *Sci. Rep.* **2018**, *8*, 3506.

(28) Jacobse, P. H.; Simonov, K. A.; Mangnus, M. J. J.; Svirskiy, G. I.; Generalov, A. V.; Vinogradov, A. S.; Sandell, A.; Mårtensson, N.; Preobrajenski, A. B.; Swart, I. One Precursor but Two Types of Graphene Nanoribbons: On-Surface Transformations of 10,10′-Dichloro-9,9′-Bianthryl on Ag(111). *J. Phys. Chem. C* **2019**, *123*, 8892–8901.

(29) Simonov, K. A.; Vinogradov, N. A.; Vinogradov, A. S.; Generalov, A. V.; Zagrebina, E. M.; Svirskiy, G. I.; Cafolla, A. A.; Carpy, T.; Cunniffe, J. P.; Taketsugu, T.; et al. From Graphene Nanoribbons on Cu(111) to Nanographene on Cu(110): Critical Role of Substrate Structure in the Bottom-Up Fabrication Strategy. *ACS Nano* **2015**, *9*, 8997–9011.

(30) Ormaza, M.; Fernández, L.; Ilyn, M.; Magana, A.; Xu, B.; Verstraete, M. J.; Gastaldo, M.; Valbuena, M. A.; Gargiani, P.; Mugarza, A.; et al. High Temperature Ferromagnetism in a $GdAg_2$ Monolayer. *Nano Lett.* **2016**, *16*, 4230–4235.

(31) Bazarnik, M.; Abadia, M.; Brede, J.; Hermanowicz, M.; Sierda, E.; Elsebach, M.; Hänke, T.; Wiesendanger, R. Atomically Resolved Magnetic Structure of a Gd-Au Surface Alloy. *Phys. Rev. B* **2019**, *99*, 174419.

(32) Ormaza, M.; Fern, L.; Lafuente, S.; Corso, M.; Schiller, F.; Xu, B.; Diakhate, M.; Verstraete, M. J.; Ortega, J. E. $LaAu_2$ and $CeAu_2$ Surface Intermetallic Compounds Grown by High-Temperature Deposition on Au(111). *Phys. Rev. B* **2013**, *88*, 125405.

(33) Xu, C.; Bao, K.; Que, Y.; Zhuang, Y.; Shao, X.; Wang, K.; Zhu, J.; Xiao, X. A Two-

Dimensional ErCu 2 Intermetallic Compound on Cu(111) with Moiré-Pattern-Modulated Electronic Structures. *Phys. Chem. Chem. Phys.* **2020**, *22*, 1693–1700.

(34) Que, Y.; Zhuang, Y.; Liu, Z.; Xu, C.; Liu, B.; Wang, K.; Du, S.; Xiao, X. Two-Dimensional Rare Earth–Gold Intermetallic Compounds on Au(111) by Surface Alloying. *J. Phys. Chem. Lett.* **2020**, *11*, 4107–4112.

(35) Abadía, M.; Ilyn, M.; Piquero-Zulaica, I.; Gargiani, P.; Rogero, C.; Ortega, J. E.; Brede, J. Polymerization of Well-Aligned Organic Nanowires on a Ferromagnetic Rare-Earth Surface Alloy. *ACS Nano* **2017**, *11*, 12392–12401.

(36) Tian, Q.; He, B.; Zhao, Y.; Wang, S.; Xiao, J.; Song, F.; Wang, Y.; Lu, Y.; Xie, H.; Huang, H.; et al. Electronic Structure Evolution at DBBA/Au(111) Interface W/O Bismuth Insertion Layer. *Synth. Met.* **2019**, *251*, 24–29.

(37) Moreno, C.; Paradinas, M.; Vilas-Varela, M.; Panighel, M.; Ceballos, G.; Peña, D.; Mugarza, A. On-Surface Synthesis of Superlattice Arrays of Ultra-Long Graphene Nanoribbons. *Chem. Commun.* **2018**, *54*, 9402–9405.

(38) Moreno, C.; Panighel, M.; Vilas-Varela, M.; Sauthier, G.; Tenorio, M.; Ceballos, G.; Peña, D.; Mugarza, A. Critical Role of Phenyl Substitution and Catalytic Substrate in the Surface-Assisted Polymerization of Dibromobianthracene Derivatives. *Chem. Mater.* **2019**, *31*, 331–341.

(39) Ma, C.; Xiao, Z.; Lu, W.; Huang, J.; Hong, K.; Bernholc, J.; Li, A. P. Step Edge-Mediated Assembly of Periodic Arrays of Long Graphene Nanoribbons on Au(111). *Chem. Commun.* **2019**, *55*, 11848–11851.

(40) Huang, L.; Pan, Y.; Pan, L.; Gao, M.; Xu, W.; Que, Y.; Zhou, H.; Wang, Y.; Du, S.; Gao, H. J. Intercalation of Metal Islands and Films at the Interface of Epitaxially Grown Graphene and Ru(0001) Surfaces. *Appl. Phys. Lett.* **2011**, *99*, 163107.

(41) Huang, L.; Xu, W. Y.; Que, Y. De; Mao, J. H.; Meng, L.; Pan, L. Da; Li, G.; Wang, Y. L.; Du, S. X.; Liu, Y. Q.; et al. Intercalation of Metals and Silicon at the Interface of Epitaxial Graphene and Its Substrates. *Chinese Phys. B* **2013**, *22*, 96803.

(42) Fei, X.; Zhang, L.; Xiao, W.; Chen, H.; Que, Y.; Liu, L.; Yang, K.; Du, S.; Gao, H. J. Structural and Electronic Properties of Pb- Intercalated Graphene on Ru(0001). *J. Phys. Chem. C* **2015**, *119*, 9839–9844.

(43) Que, Y.; Zhang, Y.; Wang, Y.; Huang, L.; Xu, W.; Tao, J.; Wu, L.; Zhu, Y.; Kim, K.; Weinl, M.; et al. Graphene-Silicon Layered Structures on Single-Crystalline Ir(111) Thin Films. *Adv. Mater. Interfaces* **2015**, *2*, 1400543.

(44) Chen, H.; Que, Y.; Tao, L.; Zhang, Y. Y.; Lin, X.; Xiao, W.; Wang, D.; Du, S.; Pantelides, S. T.; Gao, H. J. Recovery of Edge States of Graphene Nanoislands Onan Iridium Substrate by Silicon Intercalation. *Nano Res.* **2018**, *11*, 3722–3729.

(45) Ruffieux, P.; Cai, J.; Plumb, N. C.; Patthey, L.; Prezzi, D.; Ferretti, A.; Molinari, E.; Feng, X.; Müllen, K.; Pignedoli, C. A.; et al. Electronic Structure of Atomically Precise Graphene Nanoribbons. *ACS Nano* **2012**, *6*, 6930–6935.

(46) Chen, Y.-C.; de Oteyza, D. G.; Pedramrazi, Z.; Chen, C.; Fischer, F. R.; Crommie, M. F. Tuning the Band Gap of Graphene Nanoribbons Synthesized from Molecular Precursors. *ACS Nano* **2013**, *7*, 6123–6128.

(47) Söde, H.; Talirz, L.; Gröning, O.; Pignedoli, C. A.; Berger, R.; Feng, X.; Müllen, K.; Fasel, R.; Ruffieux, P. Electronic Band Dispersion of Graphene Nanoribbons via Fourier-Transformed Scanning Tunneling Spectroscopy. *Phys. Rev. B* **2015**, *91*, 045429.

(48) Yang, L.; Park, C. H.; Son, Y. W.; Cohen, M. L.; Louie, S. G. Quasiparticle Energies and Band Gaps in Graphene Nanoribbons. *Phys. Rev. Lett.* **2007**, *99*, 6–9.

(49) Braun, S.; Salaneck, W. R.; Fahlman, M. Energy-Level Alignment at Organic/Metal and Organic/Organic Interfaces. *Adv. Mater.* **2009**, *21*, 1450–1472.

(50) Otero, R.; Vázquez de Parga, A. L.; Gallego, J. M. Electronic, Structural and Chemical Effects of Charge-Transfer at Organic/Inorganic Interfaces. *Surf. Sci. Rep.* **2017**, *72*, 105–145.

(51) Binnig, G.; Rohrer, H. Scanning Tunneling Microscopy. *Surf. Sci.* **1983**, *126*, 236–244.

(52) König, T.; Simon, G. H.; Rust, H.-P.; Heyde, M. Work Function Measurements of Thin Oxide Films on Metals—MgO on Ag(001). *J. Phys. Chem. C* **2009**, *113*, 11301–11305.

(53) Xu, C.; Que, Y.; Zhuang, Y.; Liu, B.; Ma, Y.; Wang, K.; Xiao, X. Manipulating the Edge of a Two-Dimensional MgO Nanoisland. *J. Phys. Chem. C* **2019**, *123*, 19619–19624.

(54) De Renzi, V.; Rousseau, R.; Marchetto, D.; Biagi, R.; Scandolo, S.; del Pennino, U. Metal Work-Function Changes Induced by Organic Adsorbates: A Combined Experimental and Theoretical Study. *Phys. Rev. Lett.* **2005**, *95*, 046804.

(55) Yu, Y. J.; Zhao, Y.; Ryu, S.; Brus, L. E.; Kim, K. S.; Kim, P. Tuning the Graphene Work Function by Electric Field Effect. *Nano Lett.* **2009**, *9*, 3430–3434.

(56) Leung, T. C.; Kao, C. L.; Su, W. S.; Feng, Y. J.; Chan, C. T. Relationship between Surface Dipole, Work Function and Charge Transfer: Some Exceptions to an Established Rule. *Phys. Rev. B* **2003**, *68*, 195408.

(57) Liu, B.; Zhuang, Y.; Que, Y.; Xu, C.; Xiao, X. STM Study of Selenium Adsorption on Au(111) Surface. *Chinese Phys. B* **2020**, *29*, 056801.

# Supporting information

# On-Surface Synthesis of Graphene Nanoribbons on Two-Dimensional Rare Earth-Gold Intermetallic Compounds

Yande Que,[1#*] Bin Liu,[1#] Yuan Zhuang,[1] Chaoqiang Xu,[1] Kedong Wang,[2*] and Xudong Xiao[1*]

[1] *Department of Physics, the Chinese University of Hong Kong, Shatin, Hong Kong, China*

[2] *Department of Physics, Southern University of Science and Technology, Shenzhen, Guangdong 518055, P. R. China*

[#] *These authors contributed equally.*

[*] Corresponding author.

Email: ydque@phy.cuhk.edu.hk, wangkd@sustech.edu.cn, and xdxiao@phy.cuhk.edu.hk

**S1. DBBA chains on Au(111)**

As shown in **Figure S1a**, the deposition of DBBA molecules onto clean Au(111) at room temperature leads to the formation of islands of DBBA molecules arranged in long chains. The molecular chains are parallel to each other and along the herringbone of Au(111). In the zoomed-in STM image (**Figure S1b**), two protrusions are revealed for each DBBA molecule within the chains, indicating the opposite tilt of the anthracene subunits in the DBBA molecules. The black rectangle in the inset of Fig. 1(b) marks the unit cell for the molecular chain structures with periodicities of 1.06 nm and 1.97 nm, slightly larger than those of DBBA on $TbAu_2$. It should be noted that another two types of DBBA molecular chains with orientation angles of 120° respective to the one presented in **Figure S1a** owing to the three-fold symmetry of Au(111) surface (**Figure S1c**).

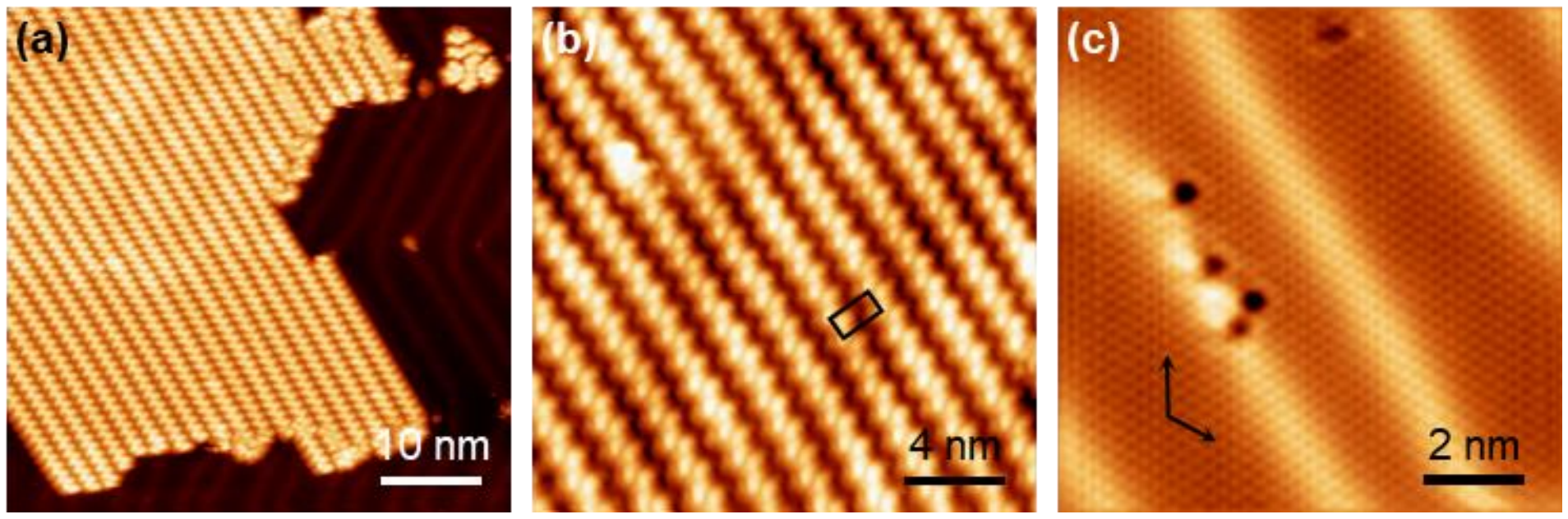

**Figure S1.** DBBA chains on Au(111). (a) Large-scale and (b) zoomed-in STM images of DBBA chains on Au(111). The black rectangle denoted the unit cell of the molecular islands. (c) Atomic-resolution STM image of clean Au(111) with a few defects on the surface. The black arrows indicate the directions of the Au(111) lattices. Tunneling conditions: (a) U = -2.0 V, I = 20 pA; (b) U = -2.0 V, I = 50 pA; (c) U = -1.0 V, I = 2.0 nA.

**S2. Additional STM image for GNRs on $TbAu_2$/Au(111) via intercalation approach**

**Figure S2** presents an additional STM image of GNRs on $TbAu_2$/Au(111) via intercalation approach, revealing the atomic lattice of $TbAu_2$ beneath the ribbons after the intercalation free of defects. Besides the ribbons with width N of 7, it also shows ribbon frags with width N of 14, formed via fusing of the neighbored ribbons. Ribbons with larger width could be easily found after annealing the sample at higher temperature, *e.g.* 450 ˚C to stimulating the fusing reaction between neighbored ribbons on Au(111). Hence, the width of the ribbons has little effect to the rare earth intercalation. In other words, other types of graphene nanoribbons on the rare earth - gold intermetallic surface alloys could also be achieved via rare earth intercalation.

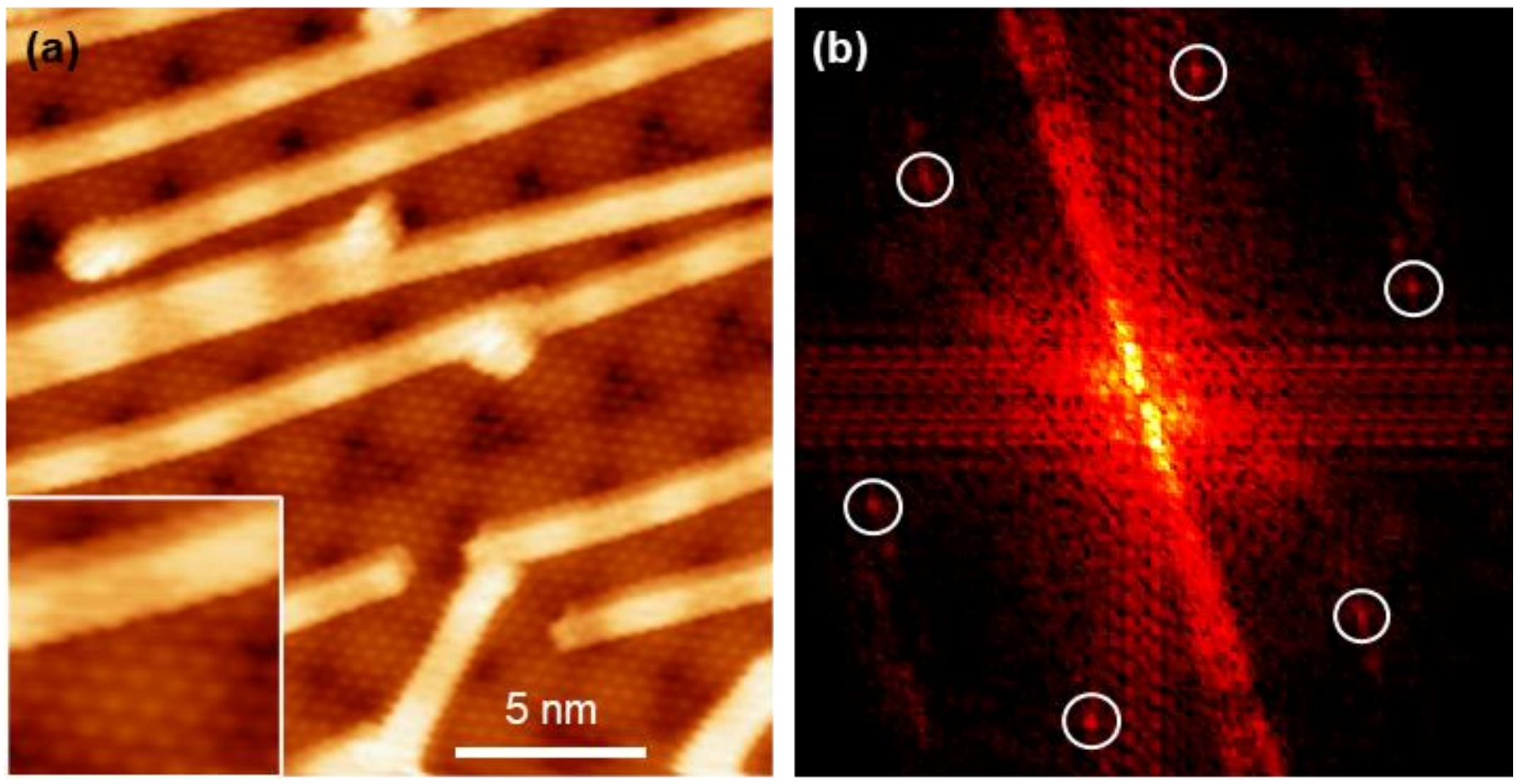

**Figure S2.** (a) Atomic-resolution STM image and (b) corresponding FFT image of GNRs/$TbAu_2$. Inset in (a) shows the zoomed-in image. The white circles in (b) denote the patterns originating from the atomic lattice of $TbAu_2$. Tunneling conditions for (a) U = -10 mV, I = 500 pA.

### S3. GNRs on $HoAu_2$/Au(111) via intercalation

As shown in **Figure S3**, deposition of Ho atoms onto GNRs/Au(111) at elevated temperature (300 ˚C) leads to the formation of well-ordered $HoAu_2$ beneath the ribbons. It should be noted that the quality of GNRs is not as high as that of GNRs on $TbAu_2$ via intercalation approach as shown in the main text. This is because the as-grown GNRs were of lower quality prior to the Ho intercalation, rather than the intercalation process affect the quality of GNRs. Nevertheless, the results demonstrate the rare earth intercalation approach could be employed for the fabrication of GNRs on other rare earth - based intermetallic compounds.

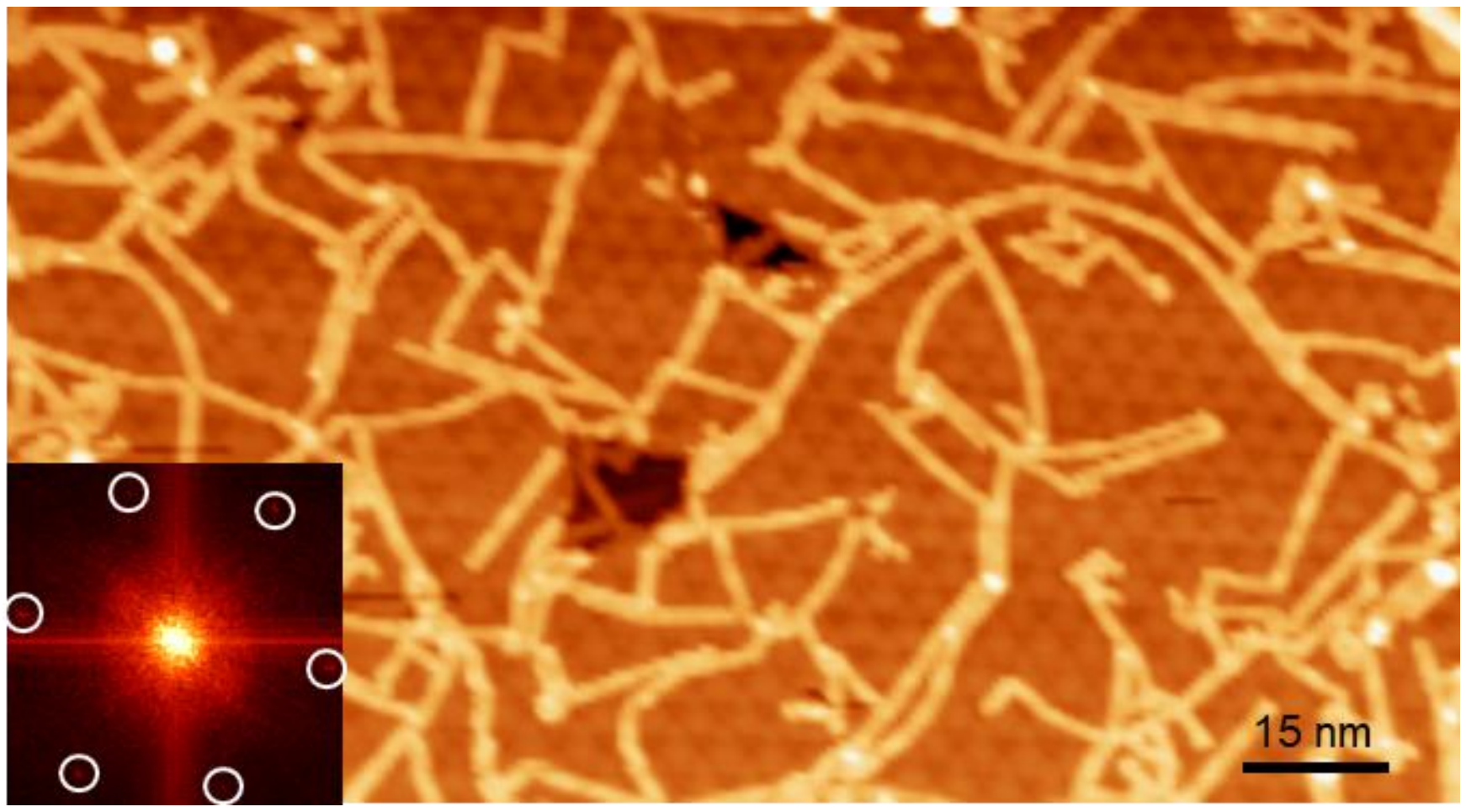

**Figure S3**. Formation of GNRs/$HoAu_2$/Au(111). Inset shows the FFT image of the STM image, where the white circles denote the patterns corresponding to the moiré structure of well-ordered $HoAu_2$. Tunneling conditions: U = -2.0 V, I = 100 pA.

## S4. Relationship between the Tunneling barrier and $I_0$

The tunneling barrier extracted from I(z) spectra might vary for different tip-sample distance range (z range). To estimate such variations, we measured I(z) spectra on clean Au(111) at different starting tunneling current $I_0$ from 0.05 nA to 0.5 nA, as shown in **Figure S4a**. The tunneling barrier slightly decreases by ~1% as the $I_0$ increases by 10 times (**Figure S4b**).

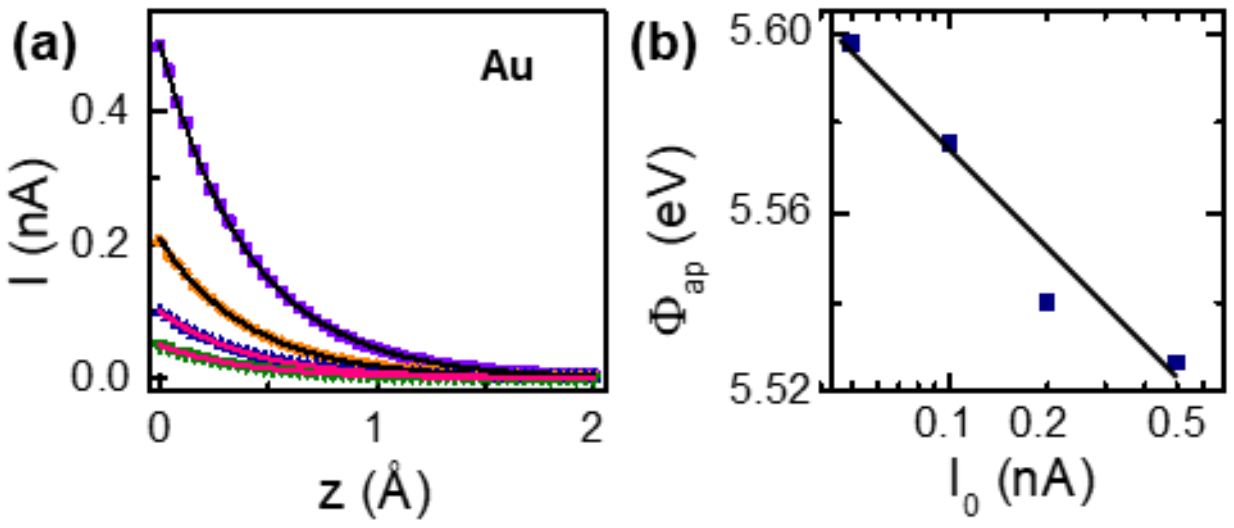

**Figure S4.** Relationship between the tunneling barrier and $I_0$. (a) I(z) spectra taken on Au(111) at $I_0$ of 0.05 nA, 0.1 nA, 0.2 nA, and 0.5 nA. the solid curves are the fittings according the equation (1) in the main text. (b) Tunneling barriers extracted from the fittings in (a). The bias for all the I(z) spectra were 2.0 V.

**S5. Unstable ribbons under FER measurement**

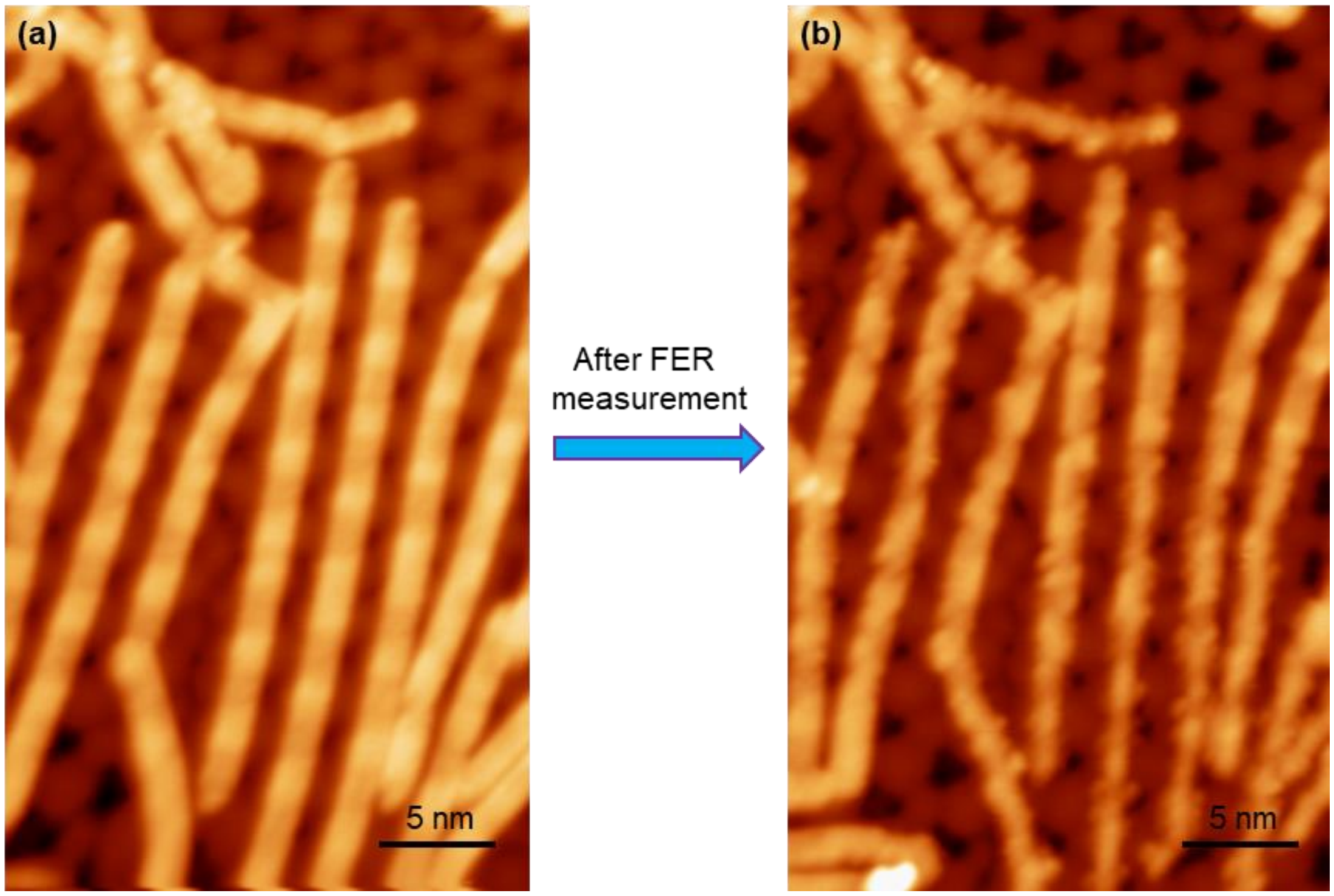

**Figure S5**. Morphology change in graphene nanoribbons after FER (dz/dV) measurement. Tunneling conditions for (a) and (b) are U = 2.0 V, I = 100 pA.

Field emission resonance could be employed to measure the surface work function locally. However, for the ribbons or some organic molecules, the high tip-induced electric field might cause the ribbons or organic molecules unstable. As shown in **Figure S5**, the morphology changed after dz/dV spectra measurement, indicating the high tip-induced electric field causes the ribbons unstable on the surface. During the dz/dV spectra measurement, the tunneling current was kept unchanged by turning on the feedback loop while swapping the sample bias from 0.5 V to 10 V. Such morphology change might lead to high uncertainty in the FER measurement.

**S6. Numerical differentiation of I(V) spectra**

Multiple I(V) spectra (typically 5 spectra) were used to be averaged to improve the sign-to-noise ratio for dI/dV spectra obtained by numerical differentiation of I(V). Numerical differentiation would effectively amplify the noise in the I(V) spectra. We used a FFT filter (window size of 11 points) to filter out the noise. The averaged I(V) spectra and corresponding dI/dV spectra are presented in **Figure S6**.

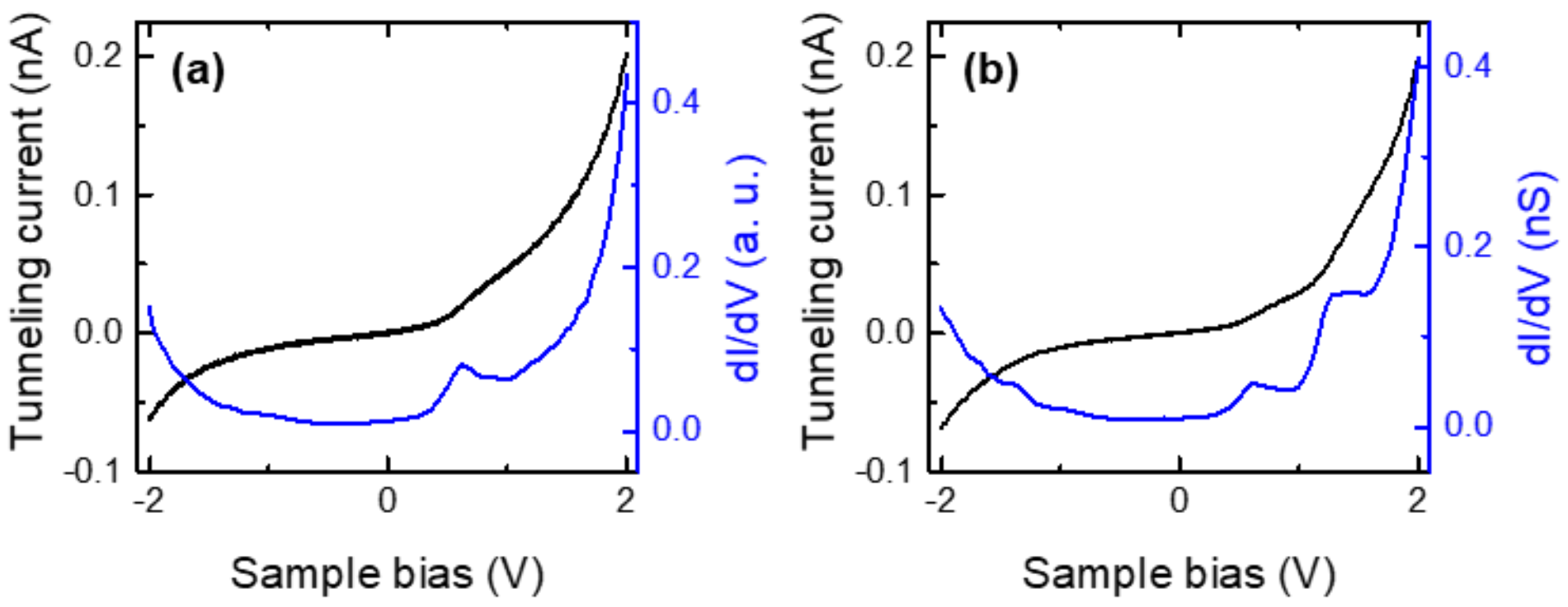

**Figure S6**. Numerical differentiation of I(V) spectra of (a) $TbAu_2$ and (b) GNR/$TbAu_2$. Tunneling conditions for (a) and (b): U = 2.0 V, I = 200 pA.